\documentclass[10pt,showpacs,twocolumn,preprintnumbers,amsmath,amssymb,
aps,prd,nofootinbib,eqsecnum,a4paper]{revtex4}
\usepackage{epsfig}
\usepackage{graphicx,epsf}
\usepackage{color}
\usepackage{bm}
\usepackage{psfrag}

\def\be{\begin{equation}}
\def\ee{\end{equation}}
\def\beb{\begin{equation*}}
\def\eeb{\end{equation*}}
\def\bea{\begin{eqnarray}}
\def\eea{\end{eqnarray}}
\def\beab{\begin{eqnarray*}}
\def\eeab{\end{eqnarray*}}
\def\nn{\nonumber}
\def \ni {\noindent}

\def\p{\partial}

\def\P{{{\cal{P}}}}

\def\H{{\cal H}}

\def\w{{\omega}}
\def\cs2{c_{\rm{s}}^2}

\def \beg {\begin{enumerate}}
\def \en {\end{enumerate}}

\def\Pb{P_0}
\def\rhob{\rho_0}

\def\drho{{\delta\rho}}
\def\dP{{\delta P}}
\def\dPn{{\delta P_{\rm{nad}}}}

\def\cs{c_{\rm{s}}^2}
\newcommand\eq[1]{Eq.~(\ref{#1})}

\def\A{\phi}

\begin{document}
\preprint{} 
\title{Vorticity generation at second order in cosmological perturbation theory}
\author{Adam J.~Christopherson$^1$}
\author{Karim A.~Malik$^1$}
\author{David R.~Matravers$^2$}
\affiliation{
$^1$Astronomy Unit, School of Mathematical Sciences, Queen Mary University
of London, Mile End Road, London, E1 4NS, United Kingdom\\
$^2$Institute of Cosmology and Gravitation, University of Portsmouth,
Dennis Sciama Building, Portsmouth, PO1 3FX, United Kingdom}
\date{\today}
\begin{abstract}
We show that at second order in cosmological perturbation theory
vorticity generation is sourced by entropy gradients. This is an
extension of Crocco's theorem to a cosmological setting.
\end{abstract}

\pacs{98.80.Cq, 98.80.Jk, 47.10.-g \hfill Phys. Rev. D 79, 123523 (2009), arXiv:0904.0940v3 }

\maketitle

\section{Introduction}

It is well known in fluid dynamics that vorticity generation is
sourced by entropy gradients. This was first pointed out by Crocco
in 1937 \cite{crocco}.
Despite vorticity being ubiquitous in nature, studies of vorticity in
the early universe and in cosmology have so far been rare.

At linear order in perturbation theory scalar and vector
perturbations, classified according to their transformation behaviour
on spatial 3-hypersurfaces, decouple from each other. Scalar
perturbations are much easier to treat mathematically and play the
dominant role in structure formation on super-horizon scales.
At linear order vorticity, intrinsically of vectorial nature (in fluid
dynamics simply the curl of the fluid velocity) cannot be constructed
from scalar quantities, and the vorticity tensor constructed from
vector perturbations is, in general, sourced only by anisotropic stress and decays in its
absence \cite{Bardeen80,KS,LLBook,Lewis:2004kg,Hollenstein:2007kg}.
\footnote{Note, however, that there are certain situations in which vorticity is not sourced
by anisotropic
stress at linear order. See Ref.~\cite{Maartens:1997sh} for an example of first order vorticity
sourced by heat flux.}

Things are different though at second order in the perturbations.
Only recently, with second order cosmological perturbation theory
becoming mature (see
e.g.~Refs.~\cite{Mukhanov:1996ak,Bruni:1996im,Noh:2004bc,MW2004,Nakamura,MM2008,MW2008}),
has the issue of vorticity generation beyond the standard linear order
begun to be addressed \cite{Matarrese:2004kq,Kobayashi:2007wd,Lu:2008ju}.
However, these previous studies have been restricted to barotropic
fluids with an equation of state $P=P(\rho)$. We show that allowing
for a non-adiabatic pressure perturbation gives qualitatively
different results.\\

We focus here on the case of vorticity generation at second order in
the perturbations, sourced only by scalar and vector perturbations.
We consider a perfect fluid, namely a fluid whose energy-momentum
tensor is diagonal (the inclusion of anisotropic stress will source
vorticity already at first order in the perturbations).  Such a fluid
has an equation of state $P=P(\rho,S)$, where $P$ is the pressure,
$\rho$ is the energy density and $S$ is the entropy of the fluid.
We can expand the pressure perturbation for a fluid with this equation
of state as
$\dP = \frac{\p P}{\p S}\delta S + \frac{\p P}{\p \rho}\drho$,
which can be written as 
$\label{defdPnad} \dP=\dPn+\cs\delta\rho\,,$ where $c_{{\rm{s}}}$ is
the adiabatic sound speed, defined as $\cs\equiv P'/\rho'$, and we
have defined the non-adiabatic pressure perturbation
$\dPn\equiv\left.\frac{\p P}{\p S}\right|_{\rho}\delta S$, in addition
to the pressure and energy density perturbations. For a detailed
discussion of the non-adiabatic pressure perturbation, see
Ref.~\cite{nonad}.  \\

Our main result, derived in detail below and given in
\eq{eq:vorsecondevolution}, can be written concisely as
\be 
\w_{2ij}^\prime\propto \drho_{1,[j}\dPn_{1,i]} \,, 
\ee 
that is, the gradients in the non-adiabatic pressure perturbation
coupled to gradients in the density act as a source for vorticity at
second order.

We use cosmological perturbation theory throughout, which will make
the numerical implementation of the results straightforward and,
compared to other approaches, has the added benefit of actually being
physically transparent. The Paper is organised as follows. In the next
section we define and introduce the key variables. In Section
\ref{sect_vort} we define the vorticity tensor and give its
evolution. We discuss our results and conclude in Section
\ref{sect_disc}. All the governing equations necessary to derive the
results in Section \ref{sect_vort} are given in the
appendix.\\

In this Paper, we use conformal time, $\eta$, throughout, denoting
derivatives with respect to conformal time with a prime. The scale
factor is $a$, and the Hubble parameter is $\H$, where
$\H=a'/a$. Greek indices, $\mu,\nu,\lambda$ run from $0,\ldots,3$,
while lower case Latin indices, $i,j,k$, take the value $1,2$, or
$3$. Covariant derivatives are denoted by a semi-colon, partial
derivatives by a comma.
The order of the perturbations is denoted with a subscript immediately
after a perturbed quantity.
We work in the uniform curvature gauge throughout.

\section{Definitions}
\label{sect_def}

The Friedmann-Robertson-Walker (FRW) metric tensor, up to and including
second order perturbations, has in the uniform curvature gauge the 
components (see, e.g.~Ref.~\cite{MW2008})
\bea
\label{metric1}
g_{00}&=&-a^2\left(1+2\A_1+\A_2\right) \,, \nn \\
g_{0i}&=&a^2\left(B_{1i}+\frac{1}{2}B_{2i}\right) \,, \nn \\
g_{ij}&=&a^2\delta_{ij}\,,
\eea
where we have assumed a flat ($K=0$) background without loss of
generality. We neglect tensor perturbations, which will add
another source term to the momentum conservation equation, but are
beyond the scope of this work \cite{inprep}.
Here $a=a(\eta)$ is the scale factor, $\A_1$ and $\A_2$ are the lapse
functions at first and second order, respectively, and $B_{1i}$ and
$B_{2i}$, represent the shear in this gauge. All perturbed quantities
are function of $x^\mu$.  Note $B_{1i}$ and $B_{2i}$ can be further
split into scalar and divergence-free vector parts
\cite{MW2008}, though this step is unnecessary here.

The fluid four velocity, $u^\mu$, obeying the normalisation condition
$u^\mu u_\mu=-1$, has components
\bea
u_0&=&-a\left[1+\A_1+\frac{1}{2}\A_2-\frac{1}{2}\A_1^2
+\frac{1}{2}v_{1k}v^k_1\right]\,,\\
u_i&=&a\left[V_{1i}+\frac{1}{2}V_{2i}-\A_1B_{1i}\right]\,,
\eea
up to second order, where $v^{i}$ is the spatial three velocity, and
$V^{i}$ is the covariant spatial velocity, defined as
$V^{i}=v^i+B^i$.

\section{Vorticity}
\label{sect_vort}

The vorticity tensor is defined as the projected anti-symmetrised
covariant derivative of the fluid four velocity, that is \cite{KS}
\be
\label{defomega}
\omega_{\mu\nu}= \P^{~\alpha}_{\mu}
\P^{~\beta}_{\nu} u_{[\alpha; \beta]}
\,,
\ee
where $u_{[\alpha; \beta]} 
\equiv \frac{1}{2}\left(u_{\alpha; \beta} - u_{\beta; \alpha}\right)$, 
and the projection tensor $\P_{\mu\nu}$ into the instantaneous fluid 
rest space is given by
\be
\label{defPmunu}
\P_{\mu\nu}=g_{\mu\nu}+u_{\mu}u_{\nu}.
\ee

The vorticity can then be decomposed, up to second
order in perturbation theory, as 
$\omega_{ij}\equiv\omega_{1ij} +\frac{1}{2}\omega_{2ij}$, where the first 
order part is simply
\be
\label{eq:omegafirst}
\omega_{1ij}=aV_{1[i,j]}\,,
\ee
and the second order part is
\be
\label{eq:omegasecond}
\w_{2ij}=aV_{2[i,j]}+2a\left[V_{1[i}^\prime V_{1j]}
+\A_{1,[i}\left(V_{1}+B_{1}\right)_{j]}
-\A_1B_{1[i,j]}\right] \,.
\ee

Using the evolution equations given in Appendix \ref{app_evol}, the
first order part evolves as
\be
\label{eq:vorfirstevolution}
\w_{1ij}^\prime - 3\H\cs\w_{1ij}=0 \,,
\ee
which gives the well known result that at first order, in the absence
of an anisotropic pressure source term, $|\w_{1ij}\w_{1}^{~ij}|\propto
a^{-2}$ during radiation domination, where $\cs=1/3$ \cite{KS}. Hence
the vorticity remains zero in this case, if it is initially zero.

However, at second order we get a non-zero source term for
the vorticity evolution equation, assuming zero anisotropic pressure.\footnote{
The calculation of the vorticity evolution equation uses first order evolution and 
field equations which we give in Appendix \ref{app_evol}.}
Even assuming zero first order vorticity,
$\w_{1ij}=0$, the second order vorticity evolves according to
\bea
\label{eq:vorsecondevolution}
&&\w_{2ij}^\prime -3\H\cs\w_{2ij}\\
&&\qquad=\frac{2a}{\rhob+\Pb}\left\{3\H V_{1[i}\dPn_{1,j]}
+\frac{\drho_{1,[j}\dPn_{1,i]}}{\rhob+\Pb}\right\}\,.\nn
\eea
Thus, we see that the non-adiabatic pressure perturbation gradients
coupled to density perturbation gradients act as a source for second
order vorticity. 
In the case of a vanishing entropy perturbation, as is
the case for barotropic fluids, we recover the result of
Ref.~\cite{Lu:2008ju}.
For completeness, we give the full second order vorticity evolution
equation without assuming $\w_{1ij}=0$ in the Appendix as
\eq{eq:vorsecondevofull}.

\section{Discussion and conclusions}
\label{sect_disc}

In this work we have shown that at second order in the perturbations
vorticity is generated from first order scalar and vector
perturbations for a perfect fluid. This is an extension of Crocco's
Theorem to an expanding, dynamical background, namely a FRW universe.

Whereas in previous works barotropic flow was
assumed, allowing for entropy gives a qualitatively novel result.
This implies that the description of the cosmic fluid as a potential
flow, which works exceptionally well at first order in the
perturbations, will break down at second order for non-barotropic
flows.
Similarly in barotropic flow Kelvin's theorem guarantees conservation
of vorticity. This is no longer the case if the flow non-barotropic.

Entropy perturbations will arise in many settings, here we
only considered the case of a single fluid with non-zero intrinsic
entropy. 
Even in simple single field inflation models, after the end of the
slow-roll regime, there is a non-zero entropy perturbation (or
non-adiabatic pressure; see, e.g., Ref.~\cite{nonad}).
Furthermore, in the case of multiple fluids we get in addition to the
intrinsic entropies of the individual fluids also an entropy component
stemming from the mixing of the fluids (the relative entropy
perturbation in the terminology of Ref.~\cite{MW2008}), even if there
is no energy and momentum transfer between the fluids.
This relative entropy perturbation can be another
source of vorticity in the multi-fluid case. However in the cosmic
plasma it is likely that the relative entropy perturbation between
fluids is subdominant compared to those mechanisms
discussed below. 
\\

Our result, that the vorticity is non-zero at second order in the
presence of non-adiabatic or entropy perturbations, has immediate
implications for the generation of magnetic fields in the early
universe, since Biermann showed that the generation of magnetic fields
is related to vorticity \cite{biermann} (see also Harrison,
Ref.~\cite{harrison}).

Previous works either used momentum exchange between
multiple fluids to generate vorticity, as in Refs.~\cite{Matarrese:2004kq, Gopal:2004ut, Takahashi:2005nd, 
Ichiki:2006cd, Siegel:2006px, Kobayashi:2007wd, Maeda:2008dv}, 
or used intermediate steps to first generate vorticity for example by using shock fronts 
as in Ref.~\cite{Ryu:2008hi}. The vorticity generated in the latter case
was then used to source turbulence. As we have shown above, we do not require
such extra steps.
We will study the evolution of magnetic fields in the case of non-zero entropy perturbations, and including tensor
perturbations, numerically in a forthcoming paper \cite{inprep}.\\

At first glance, one might expect that because the vorticity
is generated as a second order effect by products of first order density
and entropy perturbations, the vorticity will be of order $10^{-10}$. 
However, the source term is in fact constructed from gradients of the 
density and entropy
perturbations and, because of this, the effect can readily be larger 
than $\mathcal{O}(10^{-10})$.
In Fourier space, this source term will be `boosted' by the wavenumbers on small
scales. Since our mechanism will manifest itself on sub-horizon scales
and hence the wavenumbers are large, this has the potential to increase the
effect. For example, the authors of Ref.~\cite{Baumann:2007zm}
recently showed, in the context of second order tensor perturbations,
that on sub-horizon scales this effect can `boost' second order
quantities to be of comparable amplitude as first order quantities.
In fact, the WMAP five year results results \cite{WMAP5} give only a
non-zero upper bound on entropy perturbations on horizon scales. Since
our effect is on sub-horizon scales, following the argument of WMAP5
and assuming a blue spectrum for the entropy perturbation, it is likely
that this effect is non-negligible.

One possible observational signature of vorticity in the early
universe is that of B-mode polarisation of the cosmic microwave
background (CMB) radiation.  At linear order in perturbation theory,
only tensor perturbations (or gravitational waves) will produce B-mode
polarisation since scalar perturbations only produce E-modes
\cite{Spergel:1997vq}, and vector perturbations will become
subdominant during inflation, and will decay with the expansion of the
universe.  However, at second order the former is no longer true and,
as we have shown above, density perturbations can generate
vorticity. This could then produce B-mode polarisation large enough to
be observable by future CMB experiments such as Planck \cite{planck}.
\\

\acknowledgments

The authors are grateful to Kishore Ananda, Chris Clarkson and Roy Maartens
for useful discussions and comments. 
AJC is supported by the Science
and Technology Facilities Council (STFC). We used the computer algebra
package {\sc{Cadabra}} \cite{Peeters} to obtain some of the conservation 
equations, and thank Kasper Peeters for useful discussions on using the package.

\begin{appendix}
\section{}
\label{app_evol}

Here we give the evolution and constraint equations necessary to
derive the results of Section \ref{sect_vort}. For more details see
Ref.~\cite{MW2008}.

\subsection{Energy-momentum conservation}
\label{energy}

In the background the energy conservation equation is 
\be
\rho_{0}'=-3\H\left(\rho_{0}+P_{0}\right)\,.
\ee
At first order we have for the energy conservation equation 
\bea
\delta\rho_{1}'
+3\H\left( {\delta\rho_{1}}+{\delta P_{1}}\right)
&+&\left(\rho_{0}+P_{0}\right) v^k_{1,k} 
=0\,, 
\eea
and for the momentum conservation equation
\be
\label{eq:consfirst}
V_{1i}^\prime
+\H\left(1-3\cs\right) V_{1i}
+\Big[\frac{\delta P_1}{\rho_0+P_0}+\A_1\Big]_{,i} =0 \,.
\ee
At second order we only need the momentum conservation equation
\begin{widetext}
\bea
\label{momentum2}
&&\Big[\left(\rho_0+P_0\right)V_{2i}\Big]^\prime
+4\H\left(\rho_0+P_0\right)V_{2i}
+\Big[\delta P_2+ \left(\rho_0+P_0\right)\A_2
\Big]_{,i}
+2 \left(\delta\rho_1+\delta P_1\right)\A_{1,i}
+2\Big[\left(\delta\rho_1+\delta P_1\right)
V_{1i}\Big]^\prime\\
&&
+8\H\left(\delta\rho_1+\delta P_1\right)V_{1i}
-2\left(\rho_0+P_0\right)\Big[\A_1'B_{1i}+\left(\A_1^2\right)_{,i}\Big]
-2\A_1\Big[\left(\rho_0+P_0\right)V_{1i}'
+\Big( \rho_0'+P_0' +4\H\left(\rho_0+P_0\right)
\Big)V_{1i}
\Big]\nonumber\\
&&
+2\left(\rho_0+P_0\right)v_{1,~k}^kV_{1i}
+2\left(\rho_0+P_0\right)v_{1}^kV_{1i,k}
-2\left(\rho_0+P_0\right)v_{1k}B_{1,~i}^k
=0 \,. \nn
\eea
\end{widetext}

\subsection {Field equations}

At zeroth order the evolution of the scale factor is governed by the
Friedmann equation, given by
\be
\label{Friedmann}
\H^2=\frac{8\pi G}{3}a^2
\rho_0\,.
\ee
We only need the field equations up to first order in the
perturbations, a considerable simplification, since the second order
lapse function in \eq{momentum2} above cancels when calculating the
vorticity evolution.  The Einstein constraint equations at first order
are
\be
2\H B^k_{1,k}+6\H ^2\A_1=-8\pi Ga^2\drho_1 \,,
\ee
and
\be
\nabla^2B_{1i}-B^k_{1,ki}-4\H\A_{1,i}=16\pi Ga^2(\rhob+\Pb)V_{1i} \,.
\ee

\subsection{Vorticity}

Using the definition of the vorticity tensor, \eq{eq:omegasecond}
above, and the equations given in the previous subsection, we
get the evolution equation for the vorticity tensor at second order
\begin{widetext}
\bea
\label{eq:vorsecondevofull}
&&\w_{2ij}^\prime-3\H\cs\w_{2ij}
+2\left[
\left(
\frac{\dP_1+\drho_1}{\rhob+\Pb}\right)^\prime+V^k_{1,k}-X^k_{1,k}
\right]\w_{1ij}  \\
&&
+2\left(V^k_1-X^k_1\right)\w_{1ij,k}-2\left(X^k_{1,j}-V^k_{1,j}\right)\w_{1ik}
+2\left(X^k_{1,i}-V^k_{1,i}\right)\w_{1jk}\nn \\
&&\qquad=\frac{a}{\rhob+\Pb}\left\{3\H\left(V_{1i}\dPn_{1,j}-V_{1j}\dPn_{1,i}\right)
+\frac{1}{\rhob+\Pb}\left(\drho_{1,j}\dPn_{1,i}-\drho_{1,i}\dPn_{1,j}\right)\right\}\nn \,, 
\eea
\end{widetext}
\ni where $X_{1i}$ is given entirely in terms of matter perturbations as 
\be
X_{1i}=\nabla^{-2}\left[\frac{4\pi G}{\H}a^2\left(3\H(\rhob+\Pb)V_{1i}
-\drho_{1,i}\right)\right]\,.
\ee

\end{appendix}



\end{document}